\documentclass[aps,prl,superscriptaddress,showpacs,twocolumn]{revtex4}

\usepackage{amsmath,amssymb}
\usepackage{graphicx}

\begin{document}
\title{Glass former units and transport in ion-conducting network glasses}
\author{Michael Schuch}
\affiliation{Fachbereich Physik, Universit\"at Osnabr\"uck,
  Barbarastra{\ss}e 7, 49076 Osnabr\"uck, Germany}
\author{Christian Trott}
\affiliation{Institut f\"ur Physik, Technische
Universit\"at Ilmenau, 98684 Ilmenau, Germany}
\author{Philipp Maass}
\email{pmaass@uos.de}
\homepage{http://www.statphys.uni-osnabrueck.de}
\affiliation{Fachbereich Physik, Universit\"at Osnabr\"uck,
  Barbarastra{\ss}e 7, 49076 Osnabr\"uck, Germany}

\date{September 9, 2010}

\begin{abstract}
  A new theoretical approach is presented for relating structural
  information to transport properties in ion conducting network
  glasses. It relies on the consideration of the different types of
  glass forming units and the charges associated with them. Changes in
  the compositions of these units lead to a re-distribution of Coulomb
  traps for the mobile ions and to a subsequent change in long-range
  ionic mobilities. It is furthermore shown how measured changes of
  the unit compositions can be explained by thermodynamic modeling.
  The theories are tested against experiments on borophosphate glasses
  and yield good agreement with the measured data both for the
  compositional changes of the units and the variation of the
  activation energy.
\end{abstract}


\maketitle
The chemical composition of ion conducting glasses can be varied to a
large extent and this offers many possibilities to optimize these
materials with respect to different demands, in particular to high
ionic conductivities \cite{Ingram:1987}. It is therefore important to
get an understanding of the connection between the network forming
structure and the long-range ionic transport properties. Considerable
progress has been made in the past to gain insight into near and
medium range order properties of ion conducting glasses by various
experimental probes such as X-ray and neutron scattering, infrared and
Raman spectroscopy, and solid-state NMR techniques (for a review, see
\cite{Greaves/Sen:2007}). A challenge is to utilize this information
for theoretical models of the ionic transport. One promising route was
suggested some time ago by building Reverse Monte Carlo models of the
glass structure based on diffraction data \cite{Swenson/etal:2001} and
by further analyzing these structural models with the bond valence
method \cite{Swenson/Adams:2003} to explore the preferred diffusion
pathways of the mobile ions.

\begin{figure}[b!] 
\centering
 \includegraphics[width=0.35\textwidth,clip=,]{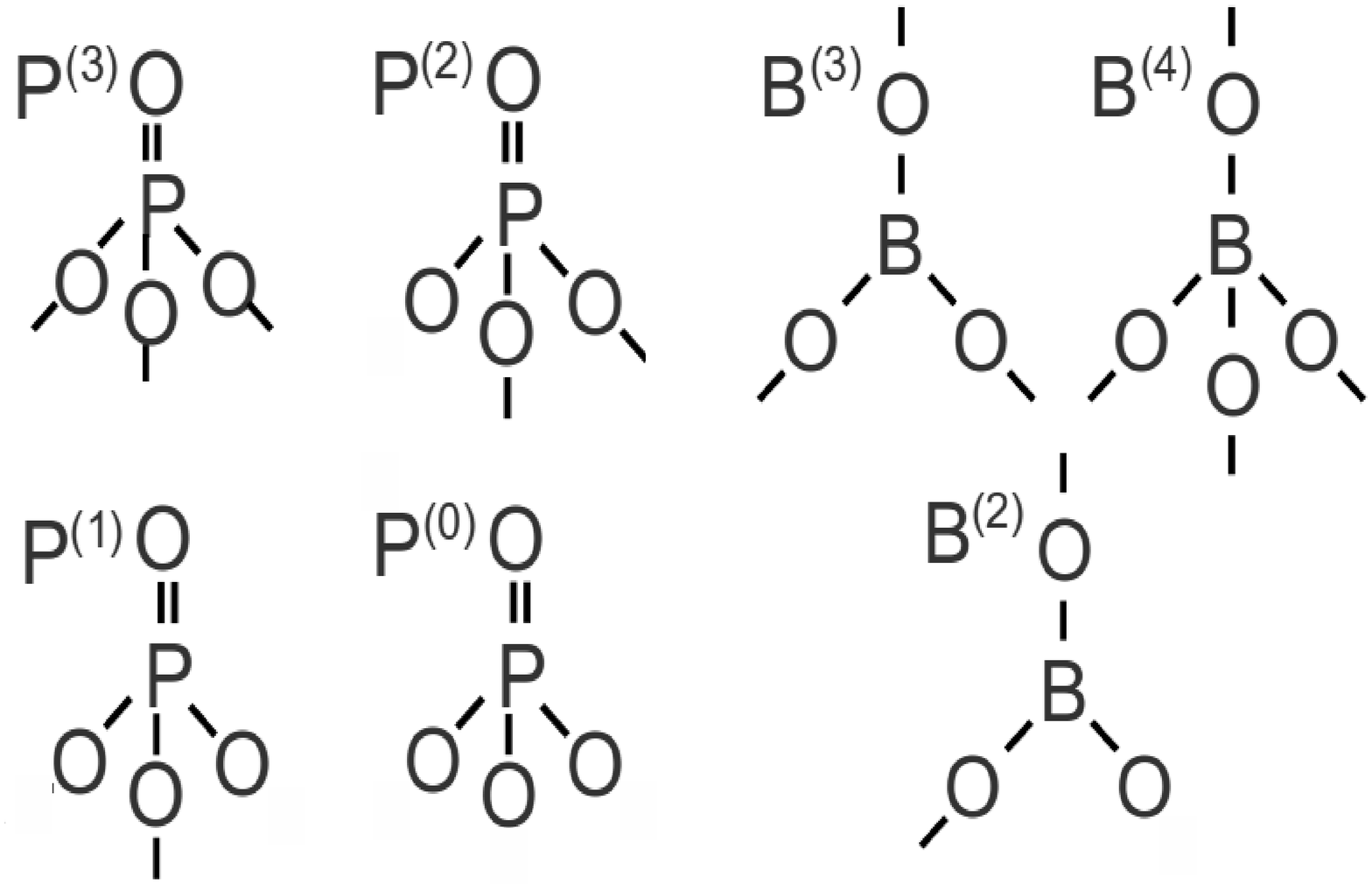}
 \caption{Sketch of the NFUs in borophosphate glasses.}
\label{fig:fig1}
\end{figure}

\begin{figure}[t!] 
\centering
 \includegraphics[width=0.4\textwidth,clip=,]{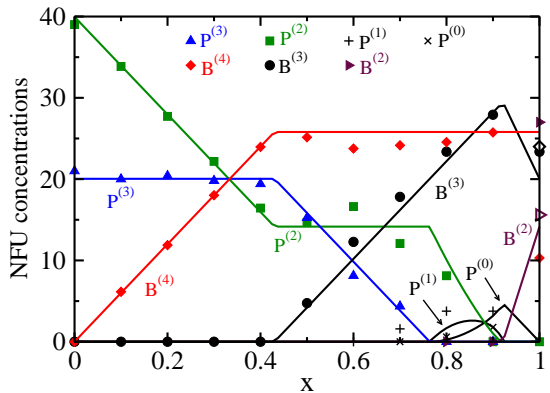}
 \caption{(color online). NFU concentrations in the glass
   0.4Na$_2$O-0.6[$x$B$_2$O$_3$-$(1-x)$P$_2$O$_5$]. The symbols mark
   the MAS-NMR results from \cite{Zielniok/etal:2007} and the open
   symbols at $x=1$ MAS-NMR measurements from
   \cite{Michaelis/etal:2007}. The solid lines are the result of the
   thermodynamic model for $\Delta\to\infty$.}
\label{fig:fig2}
\end{figure}

In this Letter we will present a new theoretical approach, which is
applicable to network forming glass structures and relies on the
different network forming units (NFUs) that build up the host
structure for the ionic motion (cf.\ Fig.~\ref{fig:fig1}). We argue
that the charges associated with the NFUs and the way how they are
localized are of crucial relevance for characterizing the statistical
properties of the energy landscape that govern the long-range ionic
transport properties. To demonstrate the new approach we apply it to
the mixed glass former effect in sodium borophosphate glasses, where detailed
information on the NFU concentrations has been gained recently by
MAS-NMR \cite{Zielniok/etal:2007,Raskar/etal:2008}, see
Fig.~\ref{fig:fig2}. We first show how the observed changes of NFU
concentrations with the borate-to-phosphate mixing ratio can be
understood from a thermodynamic model. Then we will use this
structural information on the NFUs to calculate changes of the
conductivity activation energy upon the mixing ratio.

In borophosphate glasses of composition
$y$Na$_2$O-$(1-y)$[$x$B$_2$O$_3$-$(1-x)$P$_2$O$_5$] we distinguish
seven NFUs as in \cite{Zielniok/etal:2007}: the neutral trigonal
B$^{(3)}$ units with three bridging oxygens (bOs) and zero
non-bridging (nBOs), the negatively charged tetrahedral B$^{(4)}$
units with four bOs and zero nbOs, the trigonal B$^{(2)}$ units with
two bOs and one negatively charged nbO, and the tetrahedral phosphate
units P$^{(n)}$, $n=0,\ldots3$ with $n$ bOs and $(3-n)$ nbOs, see the
Fig.~\ref{fig:fig1}. MAS-NMR measurements redrawn in
Fig.~\ref{fig:fig2} (symbols) show that, when starting the mixing from
the phosphate rich side ($x=0$), first the B$^{(4)}$ units replace
P$^{(2)}$ units. This replacement continues until the B$^{(4)}$
concentration saturates at about $x\simeq0.4$. Above this mixing
concentration the neutral B$^{(3)}$ units start to appear, replacing
now the neutral P$^{(3)}$ units to keep the total amount of negative
charge constant. This is needed to compensate the positive charge of
the mobile sodium ions. With further increasing $x$, the behavior
becomes more complex until at the boron rich side all NFUs are somehow
involved in forming the network structure.


To understand this behavior we developed a thermodynamic model, which
is based on a hierarchy of formation enthalpies $G(X)$ for the NFUs.
The B$^{(4)}$ are the most preferable NFUs for the charge compensation
of the alkali ions, since they are most highly connected in the
network with their four bOs. However, their concentration is limited,
because the delocalized charge hinders them to come close to each
other \cite{Wright:2010}. Defining in general the NFU concentrations
$[X]$ as the fraction per network former cation (i.e.\ total number of
units $X$ divided by total number of B and P atoms) we set $[{\rm
  B}^{(4)}]=\min(x,[{\rm B}^{(4)}]_{\rm sat})$, where the saturation
limit $[{\rm B}^{(4)}]_{\rm sat}=0.43$ is chosen in agreement with the
pioneering MAS-NMR results by Bray and coworkers
\cite{Bray/OKeefe:1963} and more recent findings
\cite{Michaelis/etal:2007}, as well as early theoretical modeling
\cite{Beekenkamp:1965}. For the other NFUs we choose the neutral ${\rm
  P}^{(3)}$ unit as reference point, and introduce only one parameter
$\Delta$ to describe the relative formation enthalpies
\begin{subequations}
\begin{align}
&\delta G({\rm P}^{(3)})=\delta G({\rm B}^{(3)})=\delta G({\rm P}^{(2)})=0
\label{eq:enthalpies_a}\\
&\delta G({\rm P}^{(1)})=\Delta\,,\hspace{0.5em} \delta G({\rm
  P}^{(0)})=2\Delta\,,\hspace{0.5em} \delta G({\rm B}^{(2)})=3\Delta
\label{eq:enthalpies_b}
\end{align}
\label{eq:enthalpies}
\end{subequations}
This choice expresses that the poorly connected and highly charged
${\rm P}^{(1)}$ and ${\rm P}^{(0)}$ units are increasingly less
preferable compared to the better connected ${\rm B}^{(3)}$, ${\rm
  P}^{(3)}$ and ${\rm P}^{(2)}$ units \cite{comm:p2} and that the
${\rm B}^{(2)}$ units are least likely due to their nbO and trigonal
configuration, which makes it difficult to accommodate them within the
network. In addition we need to take care of the total amount of
negative charge, being fixed by the sodium content due to charge
neutrality, and the total amount of borate and phosphate given by $x$.
This yields the constraints
\begin{subequations}
\begin{align}
&[{\rm B}^{(4)}]+[{\rm B}^{(2)}]+[{\rm P}^{(2)}]+2[{\rm P}^{(1)}]+
3[{\rm P}^{(0)}]=\frac{y}{1-y}
\label{eq:charge_constraint}\\[-1ex]
&[{\rm B}^{(4)}]+[{\rm B}^{(3)}]+[{\rm B}^{(2)}]=x
\label{eq:boron_constraint}
\\
&[{\rm P}^{(3)}]+[{\rm P}^{(2)}]+[{\rm P}^{(1)}]+[{\rm P}^{(0)}]=(1-x)
\label{eq:phosphor_constraint}
\end{align}
\label{eq:constraints}
\end{subequations}
In a grand-canonical treatment we can assign the chemical potentials
$\mu_q$, $\mu_{\rm\scriptscriptstyle B}$ and
$\mu_{\rm\scriptscriptstyle P}$ to these constraints
(\ref{eq:charge_constraint})-(\ref{eq:phosphor_constraint}), respectively.
Considering a set of sites to be occupied by the NFUs with mutual site
exclusion, we obtain the generalized Fermi distributions
\begin{subequations}
\begin{align}
&[X^{(3)}]=\frac{1}{\exp[-\mu_{\scriptscriptstyle X}]+1}\,,\quad
X={\rm B}^{(3)},{\rm P}^{(3)}
\label{eq:x3_occupation}\\
&[{\rm B}^{(2)}]=\frac{1}{\exp[3\Delta-\mu_q
-\mu_{\scriptscriptstyle B}]+1}
\label{eq:b2_occupation}
\\
&[{\rm P}^{(n)}]=\frac{1}{\exp[(2-n)(\Delta-\mu_q)
-\mu_{\scriptscriptstyle P}]+1}\,,\quad n=0,1,2
\label{eq:palpha_occupation}
\end{align}
\label{eq:nfu_occupations}
\end{subequations}
where all energies are given in units of the thermal energy $k_{\rm
  B}T$ and the chemical potentials have to be determined from
Eqs.~(\ref{eq:charge_constraint})-(\ref{eq:phosphor_constraint}).

Equations~(\ref{eq:enthalpies}) with the single parameter $\Delta$
describe the hierarchy between the formation enthalpies. Specific
values for these enthalpies should be irrelevant as long as the system
is in the low-temperature regime. To evaluate the behavior in this
regime we solve the set of
Eqs.~(\ref{eq:constraints},\ref{eq:nfu_occupations}) for
$\Delta\to\infty$. The results shown as solid lines in
Fig.~\ref{fig:fig2} are in good agreement with the MAS-NMR data from
ref.~\cite{Zielniok/etal:2007} (diamonds), except for $x=1$, where the
measured $[{\rm B}^{(4)}]$ is much smaller than the presumed
saturation value $[{\rm B}^{(4)}]_{\rm sat}=0.43$, and correspondingly
the $[{\rm B}^{(2)}]$ value larger than the theoretical prediction.
With respect to the deviation at $x=1$, we note that MAS-NMR
measurements reported by another group \cite{Michaelis/etal:2007}
yield the data marked by the open symbols in Fig.~\ref{fig:fig2}, which are
in better agreement with the theoretical predictions. On the basis of
the thermodynamic model, one can, of course, reproduce the behavior
found in ref.~\cite{Zielniok/etal:2007} by assuming a lower saturation
value $[{\rm B}^{(4)}]_{\rm sat}$ for $x=1$. Indeed, for the sodium
borate glass, the maximal $[{\rm B}^{(4)}]$ was found to be slightly
smaller than 0.43 \cite{Wright:2010}, which can be explained by
requiring that a bO cannot link two B$^{(4)}$ units
\cite{Beekenkamp:1965}. However, to describe all details, including
different behaviors for different types of alkali ions, one needs to
weaken this rule and allow for the formation of diborate groups
\cite{Kamitsos/etal:1987}. Let us note that by including such
refinements it is also possible to model the $[{\rm B}^{(4)}]_{\rm
  max}$ in the borophosphate system. To keep things simple we have
focused on the essential idea and used the limit $[{\rm B}^{(4)}]_{\rm
  max}\simeq0.43$ here.

Next we show how one can, based on the information on the NFU
concentrations, successfully model long-range ionic transport
properties. To this end we developed a model, which we call the
Network Unit Trapping (NUT) model. It relies on the following idea:
the nbOs create localized Coulomb traps for the mobile ions, while
delocalized charges, as those of the ${\rm B}^{(4)}$ units, give a
partial Coulomb contribution to several neighboring ion sites. In this
way the structural energy landscape for the ionic pathways is modified
with the mixing concentration $x$ and this effect can be conjectured to govern
the change of the activation energy $E_{\rm a}(x)$ for the long-range
ionic transport.

\begin{figure}[t!] 
\centering
 \includegraphics[width=0.45\textwidth,clip=,]{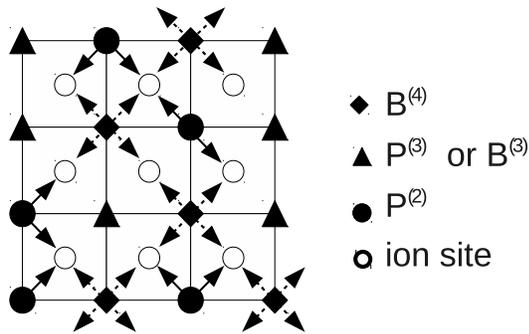}
 \caption{Two-dimensional sketch of the NUT model. The arrows indicate
   charge transfer to ion sites as described in the text.}
\label{fig:fig3}
\end{figure}

To test this model we randomly distribute the NFUs with their
concentrations from Eqs.~(\ref{eq:nfu_occupations}) on the sites of a
simple cubic lattice. These sites are called NFU sites. The mobile
ions are considered to perform a hopping motion between the centers of
the lattice cells, which represent the ion sites. An NFU $\alpha$ with
$k_\alpha>0$ nbOs and charge $(-z_\alpha e)$ adds a Coulomb
contribution $(-z_\alpha e/k_\alpha)$ to $k_\alpha$ randomly selected
neighboring ion sites, as illustrated in Fig.~\ref{fig:fig3}. Note
that this implies that the delocalization of electrons belonging to
the double bond in the charged ${\rm P}^{(n)}$ units is taken into
account. For example, a ${\rm P}^{(2)}$ unit on an NFU site $i$
induces a charge $-e/2$ at two randomly selected neighboring ion
sites. The delocalized charge of a ${\rm B}^{(4)}$ unit is spread
equally among the neighboring ion sites, which amounts to set $k=8$
for this unit. The neutral ${\rm B}^{(3)}$ and ${\rm P}^{(3)}$ units
give no Coulomb contribution. Finally, Gaussian fluctuations are added
to the site energies in order to take into account the disorder in the
glassy network \cite{comm:topology}. In summary we can write for the
energy of ion site $i$
\begin{equation}
E_i=-E_0\left[\sum_{\alpha,j}\frac{z_\alpha}{k_\alpha}\,\xi_{i,j}^\alpha+
\eta_i\right]
\label{eq:site_energies}
\end{equation}
where the sum over $j$ runs over all neighboring NFU sites of ion site
$i$. The occupation number $\xi_{i,j}^\alpha$ is equal to one, if an
NFU $\alpha$ on site $j$ contributes a Coulomb contribution $-z_\alpha
e/k_\alpha$ to ion site $i$; otherwise it is equal to zero. The
parameter $E_0>0$ sets the energy scale and the $\eta_i$ are
independent Gaussian random variables with zero mean and standard
deviation $\sigma$. Note that $E_0$ is irrelevant as long we are
interested in relative changes of the activation energy with $x$.
Hence $\sigma$ is the only tunable parameter in the modeling.

\begin{figure}[t!] 
\centering
 \includegraphics[width=0.45\textwidth,clip=,]{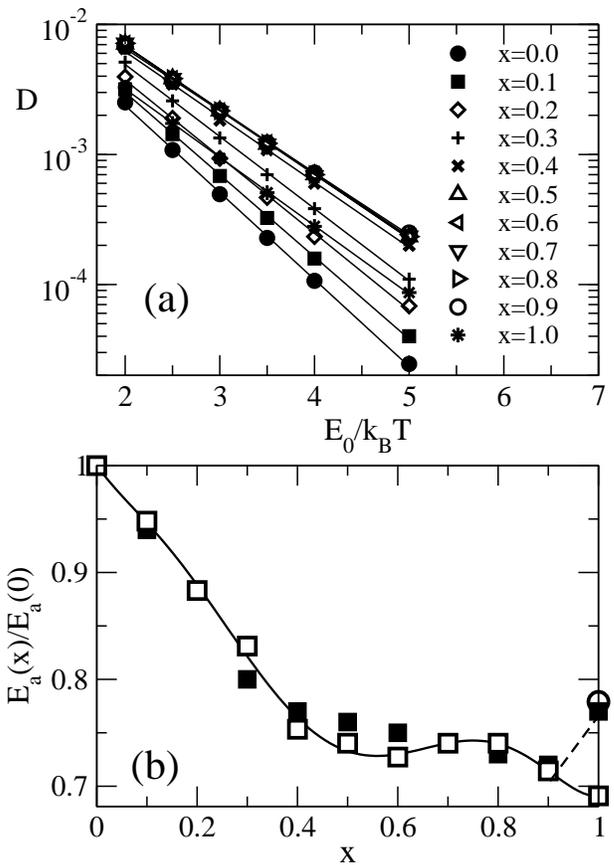}
 \caption{(a) Arrhenius plot of the simulated Na$^+$ diffusion
   coefficients $D$ in 0.4Na$_2$O-0.6[$x$B$_2$O$_3$-$(1-x)$P$_2$O$_5$]
   for various $x$ and $\sigma=0.25$. $D$ is given in units of $\nu
   a^2$, where $\nu$ is the attempt frequency of the ion jumps and $a$
   is the lattice constant (mean jump distance). The slope of the
   regression lines yields the activation energies. (b) Comparison of
   the simulated activation energy (open squares) with the measured
   conductivity activation energy from \cite{Zielniok/etal:2007} (full
   squares). The solid line is a least square fit of a polynomial of
   sixth order to the open symbols. The open circle at $x=1$
   (connected with the dashed line) corresponds to the simulated
   $E_{\rm a}$, if the NFU concentrations from
   \cite{Zielniok/etal:2007} are taken.}
\label{fig:fig4}
\end{figure}

To determine the activation energy $E_{\rm a}(x)$ we have chosen a
lattice with 50$^3$ sites, occupied all NFU sites according to the
occupation probabilities given by Eqs.~(\ref{eq:nfu_occupations}), and
the ion sites randomly with concentration $y/(1-y)$. Then Kinetic
Monte-Carlo simulations with periodic boundary conditions and
Metropolis transition rates \cite{Porto/etal:2000} were performed. After thermalization the
time-dependent mean-square displacement $R^2(t)$ of the mobile ions 
 and the diffusion coefficient $D=\lim_{t\to\infty}
R^2(t)/6t$ are determined. The diffusion coefficient is shown for $\sigma=0.25$ and
various mixing concentrations in an Arrhenius plot in
Fig.~\ref{fig:fig4}a. From the slopes of the straight lines we calculated the activation energy $E_{\rm a}(x)$, and the behavior of
the normalized activation energy $E_{\rm a}(x)/E_{\rm a}(0)$ is
compared with the experimental results from \cite{Zielniok/etal:2007}
in Fig.~\ref{fig:fig4}b. The overall agreement between the theoretical
(open symbols, solid line) and the experimental data (full symbols) is
surprisingly good. Note that we needed to fit only one parameter
$\sigma$ to achieve this agreement. A significant difference between
the theoretical and experimental curve can be seen for $x\to1$: while
the theoretical $E_a(x)$ deceases monotonously with $x$, the
experimental $E_a(x)$ finally rises for the sodium-borophosphate glass
($x=1$). Interestingly, this rise is reproduced by the NUT model
(dashed line), if instead of the NFU concentrations predicted by
Eqs.~(\ref{eq:nfu_occupations}), the NFU concentrations measured in
\cite{Zielniok/etal:2007} are used. In view of the discrepancies at
$x=1$ between experiments discussed in connection with
Fig.~\ref{fig:fig2}, this calls for a reevaluation of the activation
energy in the sodium borate system.

In summary we have presented a new approach to relate structural
information to transport properties in ion-conducting network glasses.
This approach is based on a consideration of the properties of the
different NFUs building the network structure with respect to total
charge and charge delocalization. In addition we showed how MAS-NMR
results for NFU concentrations can be understood from thermodynamic
modeling. The potential of our new approach is manifold, since one can
apply it quite generally to other network glasses with different
compositions. One immediate application, for example, could be the
investigation of glass series with varying modifier content. It is
known that the activation energy often shows a logarithmic decrease
with the concentration of mobile ions
\cite{Maass/etal:1992} and it would be important to see
whether this behavior can be captured by the NUT model.

\begin{acknowledgments}
  We would like to thank H.\ Eckert and S.\ W.\ Martin for very
  valuable discussions and gratefully acknowledge financial support of
  this work by the Deutsche Forschungsgemeinschaft in the Materials
  World Network (DFG Grant number MA~1636/3-1).
\end{acknowledgments}


\begin{thebibliography}{99}
\bibitem{comm:p2} One should expect $\delta G({\rm
    P}^{(2)})=\Delta'>0$, but this $\Delta'$ is supposed to be
  significantly smaller than $\Delta$ in Eq.~(\ref{eq:enthalpies_b}),
  since even in pure phosphate glasses P$^{(2)}$ units occur in
  combination with positively charged P$^{(4)}$ units. For the
  low-temperature limit (see below) the setting $\delta G({\rm
    P}^{(2)})=0$ is of no importance then.
\bibitem{comm:topology} The Gaussian fluctuations are needed, since,
  by using a regular arrangements of sites, we disregard the complex
  network topology. For long-range ionic motion at large length
  scales, we do not expect our results to be strongly influenced by
  the lattice structure.
\end{thebibliography}
\end{document}